\documentclass[aps,prd,twocolumn,floatfix,superscriptaddress]{revtex4-2}
\usepackage{tabularx}
\usepackage{graphicx}
\usepackage{amsmath}
\usepackage{amssymb}
\usepackage{xcolor}
\usepackage{xspace}
\usepackage{multirow}
\usepackage{hyperref}
\usepackage{subcaption}
\usepackage{cleveref}
\usepackage{orcidlink}
\usepackage{float}
\usepackage[normalem]{ulem}
\usepackage{siunitx}
\usepackage{bm}
\usepackage{booktabs}

\usepackage{etoolbox}

\newtoggle{showadded}
\newtoggle{showaddedNew}
\newtoggle{showdeleted}
\newtoggle{showDeletedBlock} 

\togglefalse{showadded} 
\togglefalse{showaddedNew} 
\togglefalse{showdeleted} 
\togglefalse{showDeletedBlock}

\newcommand{\added}[1]{%
  \iftoggle{showadded}{\textcolor{blue}{#1}}{#1}%
}

\newcommand{\deleted}[1]{%
  \iftoggle{showdeleted}{\textcolor{red}{\sout{#1}}}{}%
}

\newcommand{\deletedktr}[1]{}

\newcommand{\fitalgoname}{\texttt{GBSIEVER}\xspace}

\newcommand{\tdiamath}{{\rm A}}
\newcommand{\tdiemath}{{\rm E}}
\newcommand{\tditmath}{{\rm T}}

\newcommand{\fstat}{$\mathcal{F}$-statistic\xspace}

\newcommand{\ktredit}[1]{#1}

\newcommand{\firstpaper}{P1\xspace}

\newcommand{\fdot}{frequency derivative\xspace}

\DeclareMathOperator*{\argmax}{argmax}

\begin{document}
\title{Improving the resolution of double white dwarf systems with spaceborne gravitational wave observatories using a robust astrophysical prior}

\author{Shao-Dong Zhao\orcidlink{0000-0003-2357-3787}}
\affiliation{%
  Institute of Theoretical Physics \& Research Center of Gravitation, 
  School of Physical Science and Technology, Lanzhou University, 
  Lanzhou 730000, China
}
\affiliation{%
  Lanzhou Center for Theoretical Physics, 
  Key Laboratory of Theoretical Physics of Gansu Province, 
  Key Laboratory of Quantum Theory and Applications of MoE, 
  Gansu Provincial Research Center for Basic Disciplines of Quantum Physics, 
  Lanzhou University, Lanzhou 730000, China
}

\author{Xue-Hao Zhang\orcidlink{0000-0001-6803-8345} }
\affiliation{%
  Institute of Theoretical Physics \& Research Center of Gravitation, 
  School of Physical Science and Technology, Lanzhou University, 
  Lanzhou 730000, China
}
\affiliation{%
  Lanzhou Center for Theoretical Physics, 
  Key Laboratory of Theoretical Physics of Gansu Province, 
  Key Laboratory of Quantum Theory and Applications of MoE, 
  Gansu Provincial Research Center for Basic Disciplines of Quantum Physics, 
  Lanzhou University, Lanzhou 730000, China
}

\author{Soumya D.~Mohanty\orcidlink{0000-0002-4651-6438}}
\email{soumya.mohanty@utrgv.edu}
\affiliation{%
  Dept.~of Physics and Astronomy, 
  University of Texas Rio Grande Valley, 
  One West University Blvd., Brownsville, Texas 78520, USA
}

\author{Yu-Xiao Liu\orcidlink{0000-0002-4117-4176}}
\affiliation{%
  Institute of Theoretical Physics \& Research Center of Gravitation, 
  School of Physical Science and Technology, Lanzhou University, 
  Lanzhou 730000, China
}
\affiliation{%
  Lanzhou Center for Theoretical Physics, 
  Key Laboratory of Theoretical Physics of Gansu Province, 
  Key Laboratory of Quantum Theory and Applications of MoE, 
  Gansu Provincial Research Center for Basic Disciplines of Quantum Physics, 
  Lanzhou University, Lanzhou 730000, China
}

\begin{abstract}

Resolving the crowded population of double white dwarf (DWD) binaries in data from spaceborne gravitational wave (GW) observatories (e.g., LISA, Taiji) remains a major analysis challenge. Comparable performance on addressing this problem has been achieved with two main approaches: global fit, in which resolvable sources are estimated simultaneously from the data, and iterative, where sources are estimated one at a time and subtracted out from the data. While the latter is computationally efficient, methods developed under this approach have traditionally followed a frequentist framework that ignores astrophysical priors. This work incorporates a strong astrophysical prior, derived from the mass limits of detached white dwarfs and linking the GW signal frequency $f$ with its time derivative $\dot{f}$, into the iterative \fitalgoname{} pipeline. Applied to simulated LISA and LISA-Taiji network data, the method increases the number of confidently resolved sources by ${\approx}7.3\%$ (LISA-only) and ${\approx}14.6\%$ (network) at matched detection rate, respectively, and improves parameter estimation accuracy.

\end{abstract}
\maketitle


\section{Introduction}
Galactic Binaries (GBs) are expected to be the dominant sources of persistent gravitational wave (GW) signals for spaceborne observatories such as LISA~\cite{amaro2017laser}, Taiji~\cite{ruan2020taiji}, and TianQin~\cite{luo2016tianqin}. The principal component by number among GBs will be Double White Dwarfs (DWDs), with the remainder consisting of neutron star--white dwarf (NS--WD) and double neutron star (NS--NS) systems. This prevalence reflects the initial mass function of progenitor stars, which yields many more white dwarfs than neutron stars. Since interferometric GW detectors have poor directional sensitivity, signals from the entire GB population will additively appear in the data, requiring individual sources to be extracted via software from the stochastic foreground created by unresolved binaries as well as instrumental noise. Population synthesis models calibrated to local white dwarf surveys predict that $\mathcal{O}(10^4)$ DWD systems will be individually resolvable~\cite{Nelemans:2001hp,Nelemans:2003xp,korol2020populations,PhysRevD.102.063021,amaro2022astrophysics}, sitting above instrumental noise at frequencies $\gtrsim 3$~mHz and above the stochastic confusion foreground created by the remaining $\mathcal{O}(10^8)$ DWDs at lower frequencies. The spectral characteristics of the latter encode information about Galactic structure and binary evolution scenarios~\cite{Buscicchio2025} that complements the information provided by the resolved systems.

To stress-test gravitational-wave data analysis methods and foster their development, the LISA community has established a series of (Mock) LISA Data Challenges (MLDCs/LDCs)~\cite{MockLISADataChallengeTaskForce:2007iof,arnaud2007overview,babak2008mock,MockLISADataChallengeTaskForce:2009wir,baghi2022lisa}. Similarly, the Taiji project has released the Taiji Data Challenges (TDC and TDC~II)~\cite{ren2023taiji,du2026towards} to address realistic data analysis problems. Among the GB resolution pipelines that have demonstrated strong performance on LDC/TDC datasets~\cite{PhysRevD.84.063009,littenberg2020global,littenberg2023prototype,strub2022bayesian,strub2023accelerating,lu2022implementation} is \fitalgoname (Galactic Binary Separation by Iterative Extraction and Validation using Extended Range)~\cite{PhysRevD.104.024023,zhang2022resolving}, which implements an iterative scheme for Maximum Likelihood Estimation (MLE) and subtraction of individual binaries. For Gaussian stationary noise, the MLE reduces to the $\mathcal{F}$-statistic~\cite{krolak2004optimal} through analytical maximization of the single-source likelihood function over a subset of linear parameters. Global optimization of the $\mathcal{F}$-statistic is then performed numerically over the remaining parameters -- namely, the frequency $f$, \fdot $\dot{f}$, and sky location angles -- using Particle Swarm Optimization (PSO)~\cite{Kennedy:1995mri}.

While \fitalgoname employs an iterative approach, other pipelines have developed global fit methods where the likelihood function is extended to include tens of sources simultaneously, and all their parameters are estimated concurrently~\cite{littenberg2020global,littenberg2023prototype,strub2022bayesian,strub2023accelerating,strub2024global}, typically using Bayesian schemes such as Reverse Jump Markov Chain Monte Carlo~\cite{green1995reversible}. Both approaches have been applied to various challenge datasets and have demonstrated the ability to resolve large numbers of sources, O($10^4$) in $\sim 2$~yr data, at detection rates~\cite{littenberg2023prototype,strub2023accelerating,strub2024global} of $\sim 75\%$ to $\sim 88\%$, where the detection rate is defined as the number of confidently detected sources (correlation between the recovered and injected source $> 0.9$) relative to all detected ones. 


Astrophysical prior information can improve the performance of all GB resolution pipelines. For detached DWDs, whose orbital evolution is driven solely by gravitational radiation reaction, the chirp mass is confined to a well-defined range determined by the observed white dwarf mass distribution~\cite{kilic2007lowest,yuan2023elm} and the Chandrasekhar limit~\cite{chandrasekhar1931maximum}. This astrophysical constraint implies that $\dot{f}$ is not statistically independent of $f$ for DWDs, but instead occupies a deterministic range defined by the allowed chirp masses. Measurements of $\dot{f}$ also carry important information for the detached DWD systems: combining estimates of $f$, $\dot{f}$, and amplitude $\mathcal{A}$ allows the distance to the source to be inferred. Together with sky location parameters (ecliptic longitude $\lambda$ and latitude $\beta$), this enables the spatial position of the system to be determined~\cite{zhao2025estimating}. For semi-detached systems, $\dot{f}$ measurements can provide insight into the ongoing mass transfer process.

Like many other pipelines, and as a purely Frequentist scheme, \fitalgoname has traditionally employed independent uniform priors on $f$ and $\dot{f}$, with physical knowledge incorporated only through fixed bounds on $\dot{f}$~\cite{PhysRevD.104.024023}. This approach neglects the intrinsic physical correlation between $f$ and $\dot{f}$ described above. Consequently, when maximizing the $\mathcal{F}$-statistic, PSO explores an unnecessarily large volume of parameter space, reducing the probability of successful convergence to the global optimum and thereby limiting the pipeline's sensitivity to weaker sources. 

Measuring $\dot{f}$ accurately with space-based GW observatories is challenging due to its typically small magnitude, implying slow signal phase evolution. This makes previous likelihood-based search methods susceptible to spurious peaks at artificially inflated $\dot{f}$ values, particularly at low frequencies, leading to significant parameter estimation errors.

The adverse effects of neglecting the $f$--$\dot{f}$ correlation for detached binaries have been recognized and are beginning to be addressed in other GB resolution pipelines. In~\cite{littenberg2020global}, a non-uniform search range for $\dot{f}$ was introduced in a global fit algorithm applied to the LDC RADLER dataset, targeting frequencies around 3.98--4.02 mHz. Building on this concept, a joint prior on $f$ and $\dot{f}$ was implemented in~\cite{strub2023accelerating}, leveraging earlier work that linked these parameters via the Chandrasekhar mass limit~\cite{strub2022bayesian}. This prior was subsequently incorporated into a global fit algorithm and demonstrated on the LDC2a dataset, which contains a significantly larger number of sources~\cite{littenberg2023prototype}.

Optimal ways to integrate astrophysical priors, such as the one on $f$ and $\dot{f}$, into iterative search pipelines like \fitalgoname have not been systematically explored, and their impact on source identification and parameter estimation remains to be studied. Notably, the benefits of such physically motivated priors become increasingly important for future multi-detector networks (e.g., the LISA--Taiji network), where enhanced data quality and the ability to break degeneracies can expose the limitations of simplistic search strategies. Previous studies have demonstrated that such network configurations can significantly increase the number of resolvable sources and improve the precision of parameter estimation \cite{zhang2022resolving}.

In this paper, we find that, across the LDC1-4 dataset, the prior increases the number of confirmed detections by approximately 7.3\% for a single LISA interferometer and by 14.6\% for a LISA-Taiji network. It also reduces parameter estimation errors for $f$ and $\dot{f}$ by constraining the search volume. These improvements lead to cleaner subtraction of bright binaries, higher fidelity of low-SNR detections in previously confusion-limited regions, and higher detection rate.

The paper is organized as follows: Section~\ref{sec:GBSIEVER} reviews the \fitalgoname pipeline, including the mathematical foundation of the $\mathcal{F}$-statistic and the iterative extraction algorithm. Section~\ref{sec:motivation} establishes the astrophysical justification for the \fdot prior, showing the physical constraints on DWD systems, and describes its implementation in \fitalgoname. Section~\ref{sec:results} presents quantitative performance improvements for both single-detector and network configurations\deleted{, including robustness tests with synthetic catalogs}. Finally, Section~\ref{sec:conclusion} discusses implications and future directions for incorporating astrophysical priors into GW data analysis pipelines.

\section{The GBSIEVER pipeline}\label{sec:GBSIEVER}
This section summarizes the main elements of the \fitalgoname\ pipeline, providing a self-contained overview for the remainder of the paper. We first introduce the signal model and noise statistics of the TDI observables, including a derivation of the $\mathcal{F}$-statistic used for source detection (Section~\ref{subsec:Mathematical and parameter decomposition}). We then describe the iterative source extraction procedure, detailing the search, subtraction, and cross-validation steps that form the core of the pipeline (Section~\ref{subsec:GBSIEVER_short}). For a complete description of the pipeline, we refer the reader to Refs.~\cite{PhysRevD.104.024023,zhang2025efficient,zhang2022resolving}.

\subsection{Data model and $\mathcal{F}$-statistic}\label{subsec:Mathematical and parameter decomposition}
Space-based GW detectors like LISA use TDI ~\cite{tinto2014time} to suppress laser frequency noise, which exceeds GW signals by approximately four orders of magnitude. The fundamental idea is to combine measurements from the three arms with appropriate time delays such that laser noise cancels while GW signals add coherently. The basic Michelson combinations in TDI 1 configuration are~\cite{LDC1-man}:
\begin{align}
X(t) = &y_{1-32,32-2}+y_{231,2-2}+y_{123,-2} + y_{3-21} - \nonumber\\ &[y_{123,-2-33}+y_{3-21,-33}+y_{1-32,3} + y_{231}], \nonumber \\
Y(t) = &y_{2-13,13-3}+y_{312,3-3}+y_{231,-3} + y_{1-32} - \nonumber\\ &[y_{231,-3-11}+y_{1-32,-11}+y_{2-13,1} + y_{312}], \nonumber \\
Z(t) = &y_{3-21,21-1}+y_{123,1-1}+y_{312,-1} + y_{2-13} - \nonumber\\ &[y_{312,-1-22}+y_{2-13,-22}+y_{3-21,2} + y_{123}],
\end{align}
here, $X(t)$, $Y(t)$, and $Z(t)$ are the first-generation Michelson TDI observables, with $Y(t)$ and $Z(t)$ obtained from $X(t)$ by cyclically permuting the spacecraft indices $1\rightarrow 2\rightarrow 3\rightarrow 1$. In $y_{slr}$, $s$, $l$, and $r$ denote the sender, link, and receiver, respectively. The link index specifies the propagation path and its orientation, while the indices following the comma indicate the applied time delays. For example, $y_{231,2-2}$ represents a measurement sent from spacecraft $2$ to spacecraft $1$ along link $3$, with the indices $2-2$ indicating the preceding propagation paths and their associated time delays. In general, $y_{slr,i}(t)\equiv y_{slr}(t-L_i)$, where $L_i$ is the light-travel time associated with the directed link $i$. The arm-vector convention follows the LDC manual; in particular, $\mathbf{n}_3$ points from spacecraft $2$ to spacecraft $1$.
From Michelson XYZ channels, we can construct the noise-orthogonal TDI channels:
\begin{align}
A &= \frac{1}{\sqrt{2}}(Z - X), \nonumber \\
E &= \frac{1}{\sqrt{6}}(X - 2Y + Z), \nonumber \\
T &= \frac{1}{\sqrt{3}}(X + Y + Z).
\end{align}

The $A$ and $E$ channels contain virtually all astrophysical information with uncorrelated noise, while the $T$ channel is largely GW-insensitive below a few mHz~\cite{adams2010discriminating} and serves for diagnostics. We denote TDI time series in channel $I \in \{\tdiamath, \tdiemath, \tditmath\}$ by $\bar{s}^I(t)$, with Fourier transforms $\tilde{s}^I(f)$. The notation $\bar{\cdot}$ emphasizes time-domain quantities, while $\tilde{\cdot}$ indicates frequency-domain counterparts.

Each GB signal in the LISA band is characterized by eight parameters, which are customarily divided into two subsets: phase-evolution parameters $\boldsymbol\theta_{\rm \added{phase}} = \{f, \dot{f}, \lambda, \beta\}$ describing the frequency evolution and sky location of the source, and amplitude parameters $\{\mathcal A, \iota,\psi,\phi_0\}$ encoding the amplitude, polarization, and initial phase of the signal. The amplitude $\mathcal{A}$ is related to the chirp mass $\mathcal{M}$ and luminosity distance $d_L$ by \cite{katz2022assessing}:
\begin{equation}
\mathcal{A} = \frac{2(G\mathcal{M})^{5/3}}{c^4 d_L}(\pi f)^{2/3}.\added{\label{eq:amplitude}}
\end{equation}

The TDI response to a GB signal can be expressed as a linear combination of four basis templates:
\begin{equation}
\bar{s}^I(\boldsymbol{\theta}) = \sum_{k=1}^{4} a_k \mathbf{X}_k^I(\boldsymbol\theta_{\rm \added{phase}}),
\end{equation}
where $\mathbf{X}_k^I(\boldsymbol\theta_{\rm \added{phase}})$ are the $k$-th template components for TDI channel $I$, and ${a_k}$ are linear amplitude coefficients obtained by reparameterizing the amplitude parameters $\{\mathcal A, \iota,\psi,\phi_0\}$. The templates $\mathbf{X}_k^I$ encode the detector response to GWs with specific polarization and phase. This decomposition allows the amplitude parameters to be maximized analytically. 

The observed data in channel $I$ is additive: $\bar{y}^I = \bar{s}^I(\boldsymbol{\theta}_{\mathrm{true}}) + \bar{n}^I$, where $\bar{n}^I$ represents Gaussian noise with covariance set by $\tilde{S}_n^{I}(f)$. Under the assumption of stationary Gaussian noise, the likelihood function for a single source is given by
\begin{equation}
\mathcal{L}(\boldsymbol{\theta}) \propto \exp\left[-\frac{1}{2}\sum_I \langle \bar{y}^I - \bar{s}^I(\boldsymbol{\theta}), \bar{y}^I - \bar{s}^I(\boldsymbol{\theta}) \rangle^I\right],
\end{equation}
with the noise-weighted inner product defined as
\begin{equation}
\langle\bar{x},\bar{z}\rangle^{I} = \frac{1}{Nf_s}(\tilde{x} \cdot \tilde{S}_n^{I-1}) \cdot \tilde{z}^{\dagger}.
\end{equation}
Here, $\cdot$ denotes element-wise multiplication, $\tilde{S}_n^{I}(f)$ the one-sided noise power spectral density for channel $I$, $N$ the number of data points, $f_s$ the sampling frequency, and $\dagger$ the complex conjugate transpose.

We construct the template correlation matrix $\mathbf{W}$ and data correlation vector $\mathbf{U}$:
\begin{align}
\mathbf{W}_{ij}(\boldsymbol\theta_{\rm \added{phase}}) &= \sum_{I} \langle \mathbf{X}_i^I(\boldsymbol\theta_{\rm \added{phase}}), \mathbf{X}_j^I(\boldsymbol\theta_{\rm \added{phase}}) \rangle^{I}, \\
\mathbf{U}_i(\boldsymbol\theta_{\rm \added{phase}}) &= \sum_{I} \langle \bar{y}^I, \mathbf{X}_i^I(\boldsymbol\theta_{\rm \added{phase}}) \rangle^{I}.
\end{align}

The MLE for the linear parameters are obtained by solving $\hat{\bm{a}} = \mathbf{W}^{-1}\mathbf{U}$. Substituting back into the likelihood and taking the log yields the $\mathcal{F}$-statistic:
\begin{equation}
\mathcal{F}(\boldsymbol\theta_{\rm \added{phase}}) = \mathbf{U}^T(\boldsymbol\theta_{\rm \added{phase}}) \mathbf{W}^{-1}(\boldsymbol\theta_{\rm \added{phase}}) \mathbf{U}(\boldsymbol\theta_{\rm \added{phase}}).
\end{equation}

This represents the analytically maximized log-likelihood over the amplitude parameters, leaving only the phase-evolution parameters to be optimized numerically. The estimated phase-evolution parameters are:
\begin{equation}
\hat{\boldsymbol\theta}_{\rm \added{phase}} = \argmax_{\boldsymbol\theta_{\rm \added{phase}}} \mathcal{F}(\boldsymbol\theta_{\rm \added{phase}}).
\end{equation}
The corresponding amplitude parameter estimates are $\hat{\bm{a}} = \mathbf{W}^{-1}(\hat{\boldsymbol\theta}_{\rm \added{phase}}) \mathbf{U}(\hat{\boldsymbol\theta}_{\rm \added{phase}})$. The $\mathcal{F}$-statistic serves as the fitness function for PSO search over the four-dimensional phase-evolution parameter space $(f, \dot{f}, \lambda, \beta)$.

\subsection{Overview of \fitalgoname{}}\label{subsec:GBSIEVER_short}
A fundamental issue in detecting GBs with space-based observatories is that signals from millions of sources overlap in both time and frequency domains. Unlike ground-based detectors where signals are typically well-separated in time, LISA will observe a continuous superposition of millions of GB signals simultaneously. This necessitates algorithms capable of disentangling this dense superposition. While global fit methods attempt to model all sources simultaneously, their computational cost scales unfavorably with source count. Iterative methods like \fitalgoname offer a computationally tractable alternative by reducing the complex multi-source problem to a sequence of simpler single-source optimization tasks.

The algorithm is built on three core principles: (i) \textbf{iterative extraction}, which identifies and removes sources iteratively, treating the residual data after each subtraction as a new single-source problem; (ii) \textbf{frequency-domain localization for computational efficiency}, where searches are confined to narrow frequency bands and accelerated via undersampling techniques, thereby reducing computational cost while preserving essential signal information; and (iii) \textbf{cross-validation}, whereby consistency checks across independent searches over different parameter ranges are employed to filter out spurious detections.

At a high level, the procedure iterates over the following steps: identifying the most significant signal within a localized frequency band; estimating its parameters using the $\mathcal{F}$-statistic (defined in Section~\ref{subsec:Mathematical and parameter decomposition}); subtracting the corresponding reconstructed waveform from the data; and applying cross-validation checks to confirm the detection. This cycle repeats on the resulting residual until no significant signals remain, with false candidates being eliminated through consistency tests over broadened parameter ranges.

The following parts describe each component of this pipeline in detail.

\subsubsection{Frequency-domain partitioning and computational efficiency}
To manage computational complexity while maintaining detection sensitivity, the algorithm implements a frequency-domain partitioning strategy. The observable frequency range is divided into overlapping bands of 0.02~mHz width, with sources accepted only from a central 0.01~mHz acceptance zone within each band. This localization reduces parameter correlations and allows the search to focus on regions where a small number of sources dominate. Within each band, undersampling techniques exploit aliasing to reduce computational cost by factors of $10$--$1000$ while preserving nearly all signal information. This approach assumes the noise is approximately white over these narrow intervals, enabling efficient evaluation of noise-weighted inner products.

\subsubsection{Iterative Source Extraction}
The iterative extraction process follows a systematic protocol designed to handle source confusion and overlapping effects:
\begin{enumerate}
\item \textbf{Initialization}: Begin with the full dataset $\{\bar{y}^I\}$ containing all sources plus noise and partition it into narrow frequency bins.
\item \textbf{Single source optimization}: For iteration $m$ in each frequency bin, maximize the $\mathcal{F}$-statistic over the phase-evolution parameter space to obtain $\hat{\boldsymbol\theta}_{\rm \added{phase},m}$.
\item \textbf{Signal reconstruction}: Construct the full parameter estimate $\hat{\boldsymbol{\theta}}_m$ by combining phase-evolution and amplitude parameters.
\item \textbf{Residual update}: Update the data by subtracting the estimated source: 
   \begin{equation}
   \bar{y}_{m+1}^I = \bar{y}_m^I - \bar{s}^I(\hat{\boldsymbol{\theta}}_m).
   \end{equation}
\item \textbf{Termination rule}: Terminate when the signal-to-noise ratio of newly detected sources falls below a threshold or when a maximum number of iterations is reached.
\end{enumerate}

\subsubsection{Cross-Validation and False Positive Mitigation}
A critical component of \fitalgoname\ is its extended-range cross-validation framework. Each dataset undergoes two independent analyses:
\begin{itemize}
\item \textbf{Primary search}: Standard parameter ranges optimized for computational efficiency.
\item \textbf{Secondary search}: an extended search over a $\dot{f}$ range ($[-10^{-14},10^{-13}]$), which is 100 times broader than the primary search range ($[-10^{-16},10^{-15}]$) to probe parameter degeneracies.
\end{itemize}

\added{We distinguish three categories of sources. The classification relies on two correlation metrics, which we first define.} The correlation metric $R(\boldsymbol\theta,\boldsymbol{\theta}')$ between two sources $\boldsymbol{\theta}$ and $\boldsymbol{\theta}'$ is
\begin{equation}\label{eq:correlation}
R(\boldsymbol{\theta},\boldsymbol{\theta}') = \frac{\sum_I \langle\bar{s}^I(\boldsymbol{\theta}),\bar{s}^I(\boldsymbol{\theta}')\rangle^I}{\sqrt{\sum_I \|\bar{s}^I(\boldsymbol{\theta})\|^2 \sum_I \|\bar{s}^I(\boldsymbol{\theta}')\|^2}}.
\end{equation}
\added{For each source identified in the primary search, we then compute the maximum estimate-estimate correlation with sources in the secondary search within the same search bin:
\begin{equation}
    R_{\rm ee}(\hat{\theta}_{1,i}) = \max_j R(\hat{\theta}_{1,i},\hat{\theta}_{2,j}),
\end{equation}
where $\hat{\theta}_{1,i}$ denotes the parameter estimate of the $i$-th source in the primary search, and $\hat{\theta}_{2,j}$ that of the $j$-th source identified in the secondary search within the same search bin.}

The three categories are then defined as follows:
\begin{itemize}
\item \textbf{Identified}: sources found in the primary search.
\item \textbf{Reported}: identified sources whose cross-validation correlation $R_{\rm ee}$ exceeds frequency- and SNR-dependent thresholds.
\item \textbf{Confirmed}: reported sources whose correlation with the injected signal exceeds $0.9$.
\end{itemize}
To avoid counting the same source more than once, we apply a one-to-one association at the matching stage. For $R_{\rm ee}$, when several primary-search sources are matched to the same secondary-search source, only the pair with the highest correlation is retained. Likewise, when several reported sources are matched to the same injected source, only the reported source with the highest correlation is retained for the confirmed classification. As a result, each injected source contributes at most one confirmed source, and the confirmed-source counts are not inflated by duplicate recoveries.

We define the \textbf{detection rate} as the fraction of reported sources that are confirmed,
\begin{equation}
    \mathrm{detection\ rate} = N_{\rm confirmed}/N_{\rm reported},
\end{equation}
which is equivalent to $1-\rm false\ alarm\ rate$. We note that this quantity characterizes the purity of the reported catalog, rather than the fraction of detectable sources that are recovered.

Although the iterative procedure described above has proven effective, the extraction of individual DWD sources in dense frequency regions remains prone to spurious estimates of the  $\dot{f}$, particularly in low‑signal‑to‑noise regimes. This susceptibility arises because the pipeline typically employs an uninformative, uniform prior on $\dot{f}$, which ignores the physically narrow range expected for detached DWD systems. To mitigate this issue, we incorporate an astrophysically derived prior on $\dot{f}$, constructed from the allowed mass range of white dwarfs, with the goal of reducing false detections and enhancing the rigor of the iterative extraction chain.

\section{Astrophysical prior: construction and data}\label{sec:motivation}

This section describes the data sets and methods used to construct the astrophysical prior. We begin by summarizing the LDC catalog. We then derive the physical constraints that relate the frequency derivative to the chirp mass of double white dwarfs (Section~\ref{subsec:prior_intro}). Finally, we describe how these constraints are encoded into the \fitalgoname pipeline via a Tukey window regularization of the \fstat (Section~\ref{subsec:prior_construction}).

\subsection{Physical constraints on the \fdot}\label{subsec:prior_intro}

The LDC~\cite{baghi2022lisa} catalog employed in this work comprises two primary families of DWD systems: detached and semi-detached. Detached systems, which experience no mass transfer, are governed by gravitational radiation reaction, resulting in a positive \fdot ($\dot f>0$). In contrast, semi-detached systems undergo mass transfer, typically leading to a negative $\dot{f}$. The population synthesis code SeBa~\cite{nelemans2001detach,nelemans2001semidetach} generated the catalog used here, which assumes a Milky Way‑like spatial distribution that includes both Galactic disk and bulge components~\cite{adams2012astrophysical}. Detached systems dominate the catalog with 26,433,152 sources, while semi-detached systems number only 3,084,499.

For detached DWDs emitting gravitational waves, orbital evolution is well described by radiation reaction \cite{katz2022assessing}, yielding the first-order \fdot:
\begin{equation}
    \dot f=\frac{96}{5}\pi^{8/3}\left(\frac{G\mathcal M}{c^3}\right)^{5/3}f^{11/3}\label{eq:fdot},
\end{equation}
where $f = 2\, f_{\rm orb}$ for circular binaries, and the chirp mass $\mathcal{M}$ depends on the component masses $m_1$ and $m_2$ as
\begin{equation}\label{eq:chirpM_def}
    \mathcal{M} = \frac{(m_1m_2)^{3/5}}{(m_1+m_2)^{1/5}}.
\end{equation}
For a given $f$, the distribution of $\dot f$ is confined to a finite range. This physical constraint arises because heavier white dwarfs produce stronger gravitational radiation, leading to faster orbital evolution for a given orbital frequency.

The component masses of DWDs are constrained by both theory and observations. The lower mass limit is approximately $0.14\,M_\odot$ from observations of extremely low-mass white dwarfs~\cite{kilic2007lowest,yuan2023elm}, while the upper limit is set by the Chandrasekhar mass of about $1.4\,M_\odot$~\cite{chandrasekhar1931maximum}. To derive the corresponding chirp mass bounds, we consider the extreme cases. The minimum chirp mass occurs for $m_1 = m_2 = 0.14\,M_\odot$, which from Eq.~\eqref{eq:chirpM_def} yields:
\begin{equation}\label{eq:min_chrpM}
\mathcal{M}_{\mathrm{min}} = \frac{(0.14 \times 0.14)^{3/5}}{(0.28)^{1/5}} M_\odot \approx 0.12\,M_\odot,
\end{equation}
Conversely, the maximum chirp mass is achieved when $m_1 = m_2 = 1.4\,M_\odot$:
\begin{equation}
\mathcal{M}_{\mathrm{max}} = \frac{(1.4 \times 1.4)^{3/5}}{(2.8)^{1/5}} M_\odot \approx 1.22\,M_\odot,
\end{equation}
These bounds yield the physically allowed chirp mass range $\mathcal{M} \in [0.12, 1.22]\,M_\odot$. Substituting $\mathcal{M}_{\rm min}$ and $\mathcal{M}_{\rm max}$ into Eq.~\eqref{eq:fdot} gives the corresponding constraints on $\dot{f}$.

\begin{figure}[htb]
    \centering
    \includegraphics[width=0.45\textwidth]{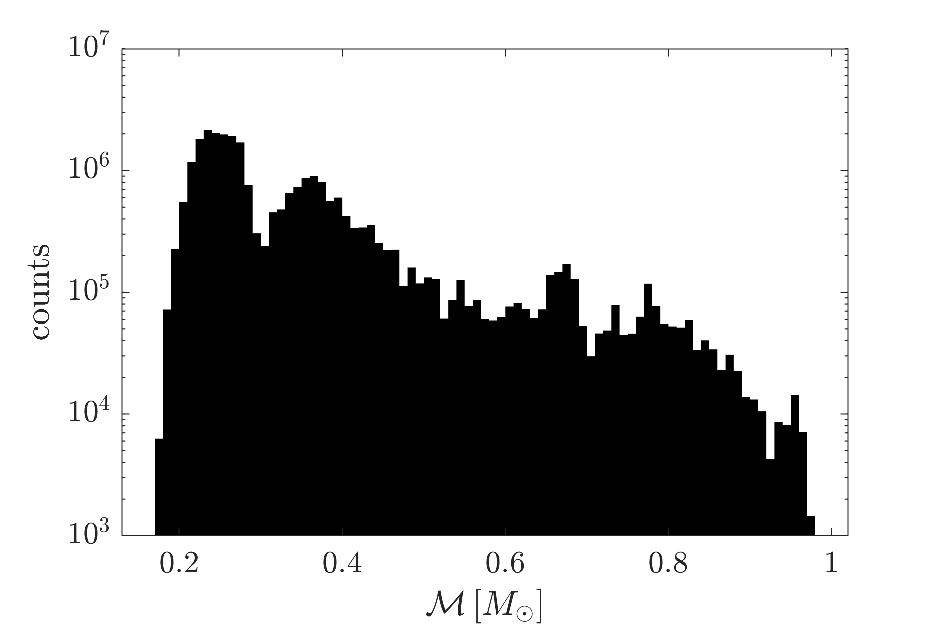}
    \caption{Chirp mass distribution of the $26,\!433,\!152$ detached double white dwarf systems in the LDC catalog. The sharply bounded distribution between approximately $\deleted{0.2}\added{0.17}\,M_\odot$ and $\deleted{1}\added{0.97}\,M_\odot$ constrains the allowed frequency derivative range and motivates the astrophysical prior introduced in Section~\ref{sec:motivation}.}
    \label{fig:chirpmass_histo}
\end{figure}

The theoretical range is derived under the assumption that the two components can independently take any mass between the observed minimum and the Chandrasekhar limit. However, the chirp mass distribution obtained directly from the GW waveform parameters of the detached systems in the LDC catalog by inverting Eq.~\eqref{eq:fdot},
\begin{equation}\label{eq:chirpM_derive}
    \mathcal{M}=\frac{c^{3}}{G}\left(\frac{5}{96} \pi^{-\frac{8}{3}} f^{-\frac{11}{3}} \dot{f}\right)^{\frac{3}{5}}\;,
\end{equation}
is much narrower, as can be seen from Fig.~\ref{fig:chirpmass_histo}. This is because the LDC detached binaries do not populate extreme configurations due to stellar evolutionary constraints: the chirp mass range actually realized by the detached branch is narrower than the theoretical one, spanning $\mathcal{M} \in [0.17, 0.97]\,M_\odot$.

For semi-detached systems, the evolution is more complex: no simple analytic relation connects $\dot f$ and the chirp mass $\mathcal{M}$. The onset of mass transfer introduces additional torques whose magnitude and sign depend on the detailed physics of mass transfer, angular momentum loss mechanisms, and the structure of the donor star~\cite{Postnov:2014tza}. Consequently, the orbital evolution of semi-detached binaries can deviate from the purely gravitational-wave-driven inspiral that characterizes detached systems. 

\begin{figure}[hbt]
    \centering
    \includegraphics[width=0.45\textwidth]{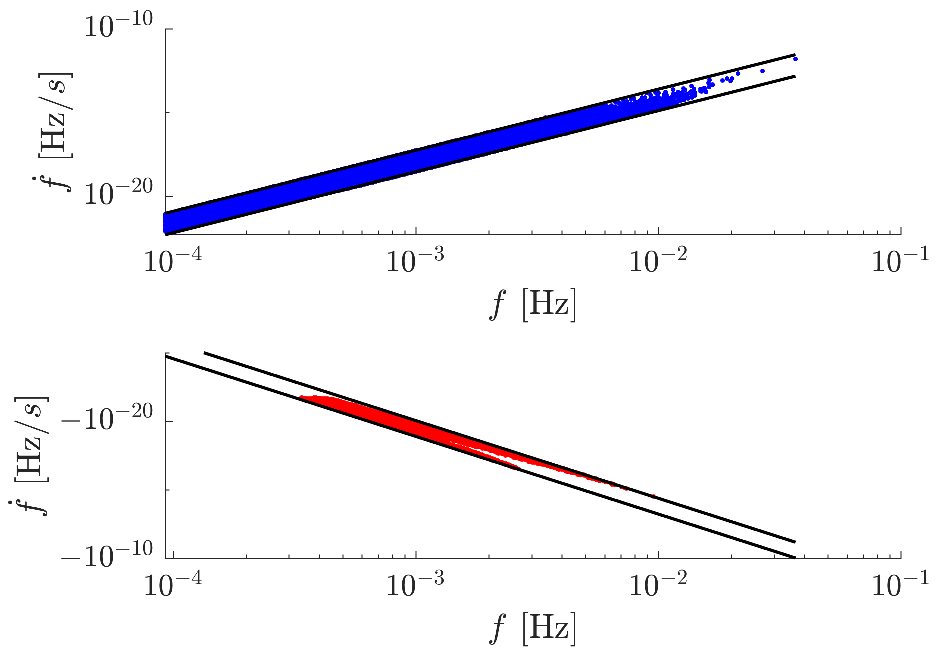}
    \caption{Frequency derivative distribution of the LDC double white dwarf catalog. Detached systems ($\dot f>0$, upper panel) and semi-detached systems ($\dot f<0$, lower panel) both occupy distinct, narrow regions of the $(f,\dot f)$ plane, limited primarily by the Chandrasekhar mass and the observed minimum white dwarf mass. This structure justifies the frequency derivative prior used in this work.}
    \label{fig:fdotConstraint}
\end{figure}

Figure~\ref{fig:fdotConstraint} shows that both the detached (positive $\dot f$) branch and semi-detached (negative $\dot f$) branches remain confined to relatively narrow bands. Despite the complicated mass transfer process, the semi-detached DWDs in the LDC catalog exhibit a well-defined structure in the $(f,\dot f)$ plane. This structured distribution suggests that, even in the absence of a simple analytic mapping, semi-detached systems can be described by priors derived from robust astrophysical modeling. At a given frequency, these bands can be characterized by four key values: the minimum and maximum allowed $\dot{f}$ for the positive branch, and likewise for the negative branch. These values are obtained from the empirical boundaries shown in Fig.~\ref{fig:fdotConstraint}, which are chosen to enclose all catalog sources.

\added{In this work, the priors are derived from the LDC catalog rather than from a specific astrophysical population model. The LDC catalog is itself generated from established population synthesis models for DWDs (e.g., \cite{nelemans2001detach,nelemans2001semidetach}), and adopting it directly avoids introducing additional model assumptions beyond those already used in the Challenge data. This choice ensures consistency with the LISA Data Challenge setup and provides a well-defined benchmark for our analysis.}


\subsection{Incorporating the prior in GBSIEVER}\label{subsec:prior_construction}

To incorporate astrophysical prior information into \fitalgoname, we reweight the $\mathcal{F}$-statistic using a smooth window along the $\dot{f}$ direction. Since $\dot{f}$ spans many orders of magnitude, from $\sim 10^{-20}$~Hz/s at low frequencies to $\sim 10^{-13}$~Hz/s at $\sim 10$~mHz, direct numerical optimization over such a wide dynamic range is computationally inefficient and susceptible to floating-point precision loss. Moreover, the physical constraint itself depends on $f$. Since PSO operates on a hypercubical search space in which the range of each parameter is specified independently, a frequency-dependent $\dot f$ bound cannot be imposed directly. To address these challenges, we introduce a dimensionless reparameterization. For each source, the sign of $\dot f$ determines which branch it belongs to, and the magnitude is normalized by the corresponding lower boundary of that branch:

\begin{equation}\label{eq:fdot_ratio}
\dot f_{\mathrm{Ratio}} =
\begin{cases}
\dfrac{\dot f}{\dot f_{\mathrm{pos,low}}(f)}, & \dot f \geq 0, \\[6pt]
\dfrac{\dot f}{\dot f_{\mathrm{neg,low}}(f)}, & \dot f < 0,
\end{cases}
\end{equation}
where $\dot f_{\mathrm{pos,low}}(f)$ and $\dot f_{\mathrm{neg,low}}(f)$ are the lower boundaries of the positive and negative branches shown in Fig.~\ref{fig:fdotConstraint}, respectively. In our implementation, these boundaries are described by power laws of the form $\log_{10}|\dot f| = a \log_{10} f + b$, with coefficients calibrated to enclose all catalog sources. \ktredit{The numerical values of the coefficients used in our analysis are listed in Table~\ref{tab:prior_coeff}}. Since $\dot f$ and its branch-specific lower boundary have the same sign, $\dot f_{\mathrm{Ratio}}$ is positive by construction, and $\dot f_{\mathrm{Ratio}} = 1$ corresponds to the lower boundary of the corresponding branch. The prior is posed on both the detached and the semi-detached branches of the DWD population. The two branches share the same window machinery described below and differ only in the upper bound of $\dot f_{\mathrm{Ratio}}$. The corresponding maximum-to-minimum ratios are $\sim 19.05$ for the detached branch and $\sim 14.13$ for the semi-detached branch. We note that the first empirical ratio is slightly larger than the chirp-mass-based ratio of $\sim 18.22$ obtained from the LDC detached range $[0.17,0.97]\, M_{\odot}$, because the drawn boundaries include a small margin beyond the extreme sources.
\begin{table}[htbp]
\caption{\ktredit{Coefficients $a$ and $b$ of the empirical power-law boundaries used to construct the lower and upper boundaries of the detached ($\dot f>0$) and semi-detached ($\dot f<0$) branches (c.f.,  Fig.~\ref{fig:fdotConstraint}). }}
\label{tab:prior_coeff}
\begin{tabular}{lcccc}
\toprule
 & \multicolumn{2}{c}{$\dot f>0$ (detached)} & \multicolumn{2}{c}{$\dot f<0$ (semi-detached)}\\
\cmidrule(lr){2-3}\cmidrule(lr){4-5}
 & lower & upper & lower & upper\\
\midrule
$a$ & $3.6399$ & $3.6399$ & $5.6581$ & $5.6581$ \\
$b$ & $-7.6000$ & $-6.3200$ & $-3.0800$ & $-1.9300$ \\
\bottomrule
\end{tabular}
\end{table}

\begin{figure}
    \centering
    \subfloat[Detached DWDs]{
        \includegraphics[width=0.45\textwidth]{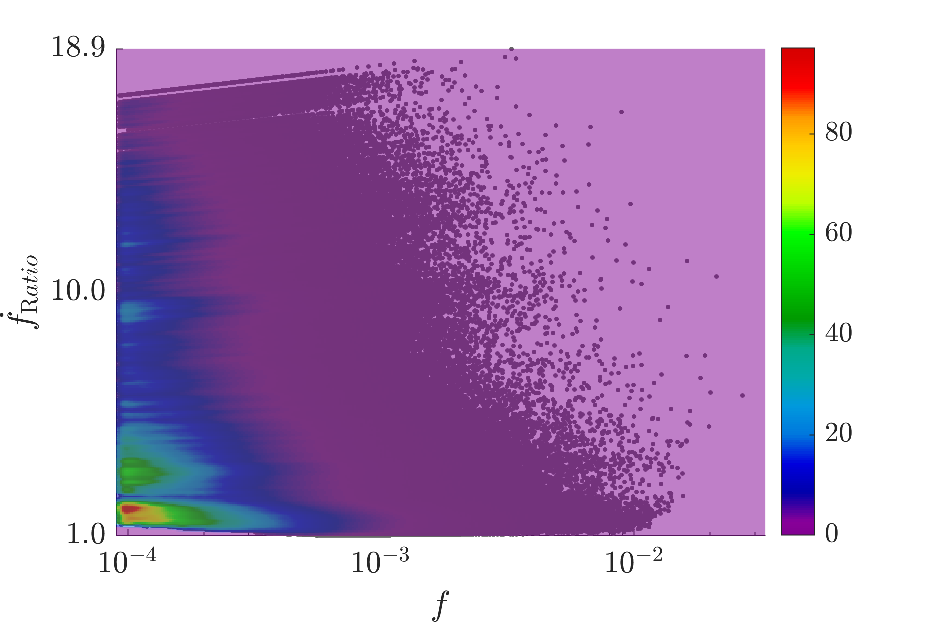}
        \label{fig:fdotR_pos}
    }
    \hfill  
    \subfloat[Semi-detached DWDs]{
        \includegraphics[width=0.45\textwidth]{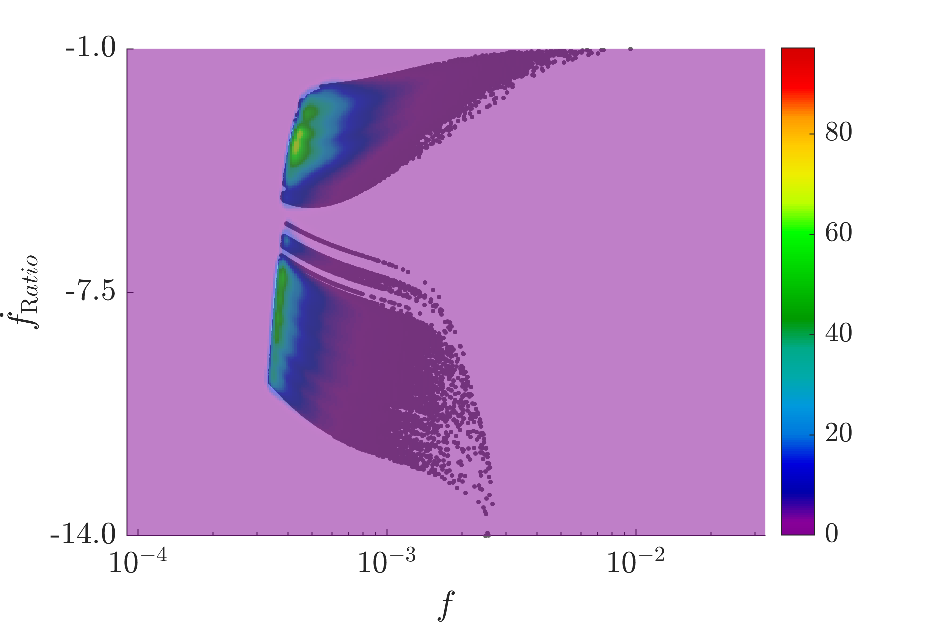}
        \label{fig:fdotR_neg}
    }
    \caption{Kernel density estimation (KDE) maps of sources with positive (top) and negative (bottom) frequency derivatives, shown in terms of the dimensionless ratio $\dot f_{\mathrm{Ratio}}$. The KDE distributions are overlaid with scatter plots of the individual sources in the parameter space. Both panels share the same color scale to allow direct visual comparison; consequently, the green shading present in the densest regions of the negative-$\dot f$ population reflects the common normalization rather than a lower intrinsic density.}
    \label{fig:fdotR_color}
\end{figure}

As shown in Fig.~\ref{fig:fdotR_color}, when plotted in terms of the dimensionless ratio $\dot f_{\mathrm{Ratio}}$, the detached and semi-detached populations each occupy a well-defined and approximately frequency-independent y-axis range. This range is the range spanned by the sources in each population along the vertical axis; no explicit boundary lines are drawn. The upper and lower edges of these y-axis ranges define the allowed $\dot f_{\mathrm{Ratio}}$ bounds used in our prior. Two distinct sub-structures in the negative $\dot f$ sources are apparent, corresponding to systems with white dwarf donors and helium star donors, respectively---a separation consistent with theoretical expectations and previous population studies~\cite{nelemans2001semidetach}. The clear boundaries make it a convenient choice for incorporating prior information via a window function. We choose a Tukey window~\cite{harris1978use}, defined as:
\begin{equation}\label{eq:tukey_window}
w(x) =
\begin{cases}
0, & x < 0 \\
\displaystyle \frac{1}{2} \left[ 1-\cos\left(\frac{\pi x}{r}\right) \right], & 0 \leq x < r \\
1, & r \leq x \leq 1-r \\
\displaystyle \frac{1}{2} \left[ 1-\cos\left(\frac{\pi (1-x)}{r}\right) \right], & 1-r < x \leq 1 \\
0, & x > 1
\end{cases}
\end{equation}
where $r = (1-\alpha)/2$ and $\alpha \in [0,1]$ controls the fraction of the window with unit weight.

The argument $x$ is the frequency derivative mapped onto the window domain. We linearly map the prior $\dot f_{\mathrm{Ratio}}$ range $[\dot f_{\mathrm{Ratio}}^{\min}, \dot f_{\mathrm{Ratio}}^{\max}]$ onto the flat central region of the Tukey window, $[(1-\alpha)/2, (1+\alpha)/2]$, through
\begin{equation}\label{eq:fdot_mapping}
x = \frac{1-\alpha}{2} + \alpha\,\frac{\dot f_{\mathrm{Ratio}} - \dot f_{\mathrm{Ratio}}^{\min}}{\dot f_{\mathrm{Ratio}}^{\max} - \dot f_{\mathrm{Ratio}}^{\min}},
\end{equation}
where $\dot f_{\mathrm{Ratio}}^{\min} = 1$, corresponding to the lower boundary defined above, and $\dot f_{\mathrm{Ratio}}^{\max}$ is specified separately for the detached and semi-detached branches, as described at the end of this subsection.

This mapping ensures that values within the physically plausible range receive unit weight ($w=1$), minimally biasing the optimization, while values outside the physical range are mapped beyond the flat region and receive smoothly tapered weights that suppress unphysical values. Values far outside the entire window domain receive the minimal Tukey weight $w_{\min}\approx0$, effectively implementing a cutoff. The parameter $\alpha=0.9$ defines the flat portion as $90\%$ of the full window.

Although the analytic Tukey window reaches exactly zero at the domain boundaries, in our implementation a small positive floor ($\sim 10^{-9}$) is imposed to avoid numerical infinities in the PSO fitness evaluation. This approximation has a negligible effect on the search outcome while ensuring numerical stability.

The fitness function introduced in Sec.~\ref{subsec:Mathematical and parameter decomposition} is then modified by the Tukey window prior:
\begin{equation}\label{eq:f_prior}
\mathcal{F}_{\rm prior}(f,\dot f,\lambda,\beta) = \mathcal{F}(f,\dot f,\lambda,\beta) \times w(x; \alpha),
\end{equation}
\added{where $x$ is the mapped frequency derivative from Eq.~\eqref{eq:fdot_mapping}, }and $w(x; \alpha)$ is the Tukey window function from Eq.~\eqref{eq:tukey_window} with smoothing parameter $\alpha=0.9$. The Tukey window serves two purposes. First, it suppresses a spurious rise in the fitness function beyond the prior range, which would otherwise attract particles outside the intended range. Second, by using smooth transitions rather than a hard cutoff, it avoids introducing a discontinuity in the fitness function that could destabilize the PSO search near the prior boundaries.

\begin{figure}[htb]
    \centering
    \includegraphics[width=0.475\textwidth]{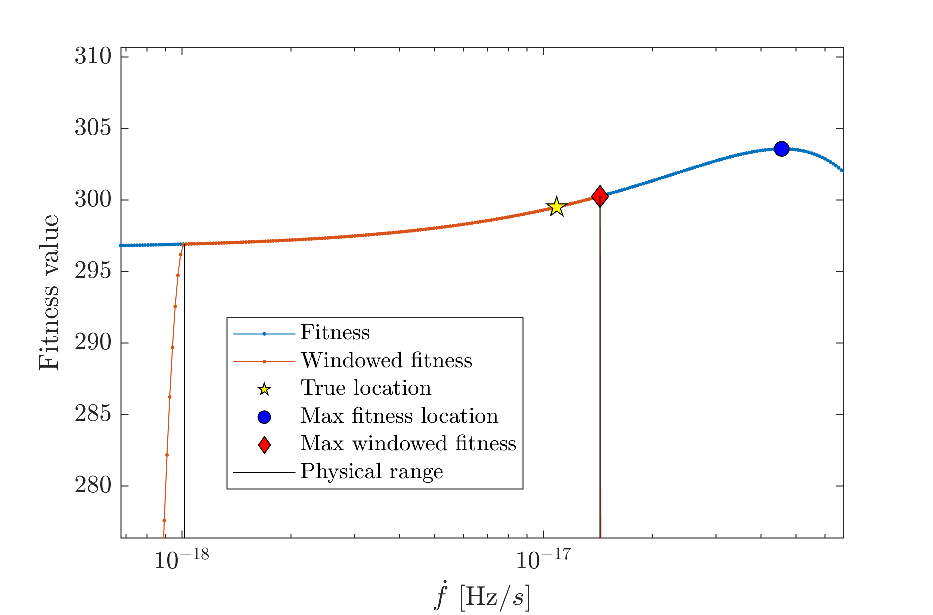}
    \caption{Impact of the Tukey-window prior on the fitness function for frequency bin 222 ($f\in[2.30,2.32]$~mHz). The blue curve represents the original $\mathcal{F}$-statistic as a function of $\dot{f}$, while the red curve shows the corresponding windowed fitness after applying the prior. The vertical lines indicate the physically allowed range of $\dot{f}$. Applying the prior suppresses the unphysical maximum outside this range and shifts the dominant peak toward the true source location, thereby improving the physical consistency of the recovered solution.}
    \label{fig:fitness_prior_cmp_222}
\end{figure}

Figure~\ref{fig:fitness_prior_cmp_222} illustrates how the prior reshapes the fitness landscape for an individual source in a representative frequency bin ($f\in[2.30,2.32]$~mHz). For this demonstration, we process the LDC data as follows: First, we remove the nine sources with ${\rm SNR}>20$ from the catalog, assuming perfect subtraction with negligible residual errors. This leaves one target source, the tenth brightest in the bin with SNR $=\added{19.12}\deleted{13}$, embedded in the residual confusion noise from the remaining $265$ fainter sources plus instrumental noise. The target's true $\dot{f}$ value falls within the astrophysically allowed range. We then construct the fitness function by fixing all target parameters at their true values, except $\dot{f}$, which we vary across the search range. In the original (unweighted) fitness landscape, the maximum occurs at a $\dot{f}$ value outside the physical range, a spurious peak induced by confusion noise that could misdirect the optimizer. Applying the Tukey window prior ($\alpha=0.9$) suppresses this unphysical maximum while preserving the correct peak within the astrophysical bounds, thereby focusing the search on the physically plausible region.

\section{Results}\label{sec:results}

We evaluate the impact of the \fdot prior using two metrics--detection efficiency, parameter estimation accuracy--across the five frequency-SNR blocks defined in Table~\ref{tab:blocks_def}. These blocks partition the LDC parameter space following the classification established in earlier GBSIEVER publications, from confusion-dominated low frequencies ($f < 3\ \rm mHz$) to the high-frequency tail where confusion is negligible. Section~\ref{subsec:LISA_results} presents single-detector results, Section~\ref{subsec:LISA-Taiji_results} covers the LISA-Taiji network.
\begin{table}[htbp]
\centering
\caption{Definition of frequency-SNR analysis blocks.}
\label{tab:blocks_def}
\begin{tabular}{c|c|c}
\toprule
 & \textbf{Frequency range (mHz)}  & \textbf{SNR range} \\
\midrule
block~1 & $[0, 3]$ & $[0, 25]$ \\
block~2 & $[0, 3]$ & $[25, \infty)$  \\
block~3 & $[3, 4]$ & $[0, 20]$  \\
block~4 & $[3, 4]$ & $[20, \infty)$ \\
block~5 & $[4, 15]$ & $[10, \infty)$ \\
\bottomrule
\end{tabular}
\end{table}

\subsection{Single LISA detector performance}\label{subsec:LISA_results}

We first assess the prior's impact on the single LISA detector configuration, which serves as the baseline for all subsequent comparisons. In Paper~1 (\firstpaper)\cite{PhysRevD.104.024023}, multiple $R_{\rm ee}$ threshold configurations were explored. Following the \textit{Main} configuration from \firstpaper, we set $R_{\rm ee}$ thresholds of $\{0.9, 0.5, 0.9, 0.5, -1\}$ for blocks 1--5 respectively, where $R_{\rm ee} = -1$ indicates that all identified sources in block~5 are treated as reported. 

Table~\ref{tab:snr_ree_LISA} summarizes the baseline performance for the single LISA detector on the Radler (LDC1-4) data using the {\it Main} set of $R_{\rm ee}$ values tuned in \firstpaper, while Table~\ref{tab:snr_ree_LISA_prior} shows the response of the pipeline once the \fdot prior is implemented. The baseline results in Table~\ref{tab:snr_ree_LISA} exhibit the expected pattern for confusion-limited searches: block~1 (low frequency, low SNR) achieves only $63.61\%$ detection rate, dragging the overall rate to $84.79\%$, while blocks~4-5 exceed $92\%$. This ~30 percentage-point gap between the worst and best blocks reflects the severity of Galactic confusion below 3 mHz.
\begin{table*}
\begin{center}
    \begin{tabular*}{0.8\textwidth}{@{\extracolsep{\fill}}ccccccc}
    \toprule
     & block~1 & block~2 & block~3 & block~4 & block~5 & Overall\\
    \midrule
    $R_{\rm ee}$ & $0.9$ & $0.5$ & $0.9$ & $0.5$ & $-1$ & -\\
    \midrule
    Identified   &  $23231$ &  $2106$ & $3696$ &  $1526$ &  $4279$ & $34838$\\
    \midrule
    Reported     &  $2767$ &  $2073$ & $1622$ &  $1510$ &  $4279$ & $12251$\\
    \midrule
    Confirmed     &  $1760$ &  $1892$ & $1303$ &  $1394$ &  $4039$ & $10388$\\
    \midrule
    Detection rate  &  $63.61\%$ &  $91.27\%$ & $80.33\%$ &  $92.32\%$ &  $94.39\%$ & $84.79\%$\\
    \bottomrule
    \end{tabular*}

\caption{Performance of the single-detector LISA implementation of \fitalgoname{} on the LDC1-4 dataset without applying the $\dot{f}$ prior. The results are grouped into the five contiguous frequency--SNR blocks used in the cross-validation analysis. The corresponding $R_{\rm ee}$ thresholds and block definitions follow the {\it Main} configuration described in Table~I of \firstpaper. The table includes both detached and semi-detached systems, classified by the sign of the estimated $\dot f$; see the main text for details.}
\label{tab:snr_ree_LISA}
\end{center}
\end{table*}
\begin{table*}
\begin{center}
\begin{tabular*}{0.8\textwidth}{@{\extracolsep{\fill}}ccccccc}
\toprule
 & block~1 & block~2 & block~3 & block~4 & block~5 & Overall\\
\midrule
$R_{\rm ee}$ & $0.9$ & $0.5$ & $0.9$ & $0.5$ & $-1$ & -\\
\midrule
Identified   & $22584$ & $2095$ & $3664$ & $1526$ & $4335$ & $34204$\\
\midrule
Reported     & $2622$ & $2058$ & $1592$ & $1510$ & $4335$ & $12117$\\
\midrule
Confirmed    & $1825$ & $1916$ & $1315$ & $1400$ & $4141$ & 10597\\
\midrule
Detection rate & $69.60\%$ & $93.10\%$ & $82.60\%$ & $92.71\%$ & $95.52\%$ & $87.46\%$\\
\midrule
\midrule
\multicolumn{7}{c}{Changing $R_{\rm ee}$ for same detection rate}\\
\midrule
\midrule
 & block~1 & block~2 & block~3 & block~4 & block~5 & Overall\\
\midrule
$R_{\rm ee}$ & $0.425$ & $0.1$ & $0.3$ & $0.1$ & $-1$ & -\\
\midrule
Identified   & $22584$ & $2095$ & $3664$ & $1526$ & $4335$ & $34204$\\
\midrule
Reported     & $3324$ & $2093$ & $1876$ & $1526$ & $4335$ & $13154$\\
\midrule
Confirmed    & $2163$ & $1934$ & $1500$ & $1410$ & $4141$ & $11148$\\
\midrule
Detection rate & $65.07\%$ & $92.40\%$ & $79.96\%$ & $92.40\%$ & $95.52\%$ & $84.75\%$\\
\bottomrule
\end{tabular*}

\caption{Performance of the single-detector LISA implementation of \fitalgoname\ on the LDC1-4 dataset after applying the $\dot{f}$ prior. The upper section presents the results obtained using the {\it Main} $R_{\rm ee}$ configuration from \firstpaper, showing improved detection efficiency and a larger number of confirmed sources compared to the baseline case without the prior (Table~\ref{tab:snr_ree_LISA}). The lower section illustrates that, by retuning the $R_{\rm ee}$ thresholds, the overall detection rate can be brought close to the baseline value while still yielding a modest increase in the number of confirmed sources. See the caption of Table~\ref{tab:snr_ree_LISA} and the main text for the source-class assignment rule.}
\label{tab:snr_ree_LISA_prior}
\end{center}
\end{table*}

The most significant improvement occurs in block~1 (low-frequency, low-SNR regime), where the detection rate increases from $63.61\%$ to $69.60\%$ under the $\dot{f}$ prior. This $5.99$ percentage points gain is accompanied by a cleaner catalog: reported candidates decrease from $2767$ to $2622$, while confirmed sources increase from $1760$ to $1825$. The reduction in false associations indicates that the prior effectively suppresses unphysical spurious peaks. The block~3 shows a smaller but consistent improvement (from $80.33\%$ to $82.60\%$), while the high-frequency block remains nearly saturated, increasing only slightly from $94.39\%$ to $95.52\%$. These trends indicate that the prior is most effective in confusion-dominated regimes where degeneracies in $\dot{f}$ previously increased false associations and reduced catalog purity.

The prior reduces the number of reported candidates by $1.1\%$ while increasing confirmed sources by $209$ ($2.0\%$ relative gain) and raising the overall detection rate from $84.79\%$ to $87.46\%$ with the {\it Main} configuration. The $R_{\rm ee}$ thresholds are retuned from their baseline values of $[0.9,0.5,0.9,0.5]$ to $[0.425,0.1,0.3,0.1]$ for LISA and $[0.425,0.1,0.5,0.1]$ for LISA-Taiji so as to match the baseline overall detection rate. The astrophysical prior yields $903$ additional reported sources and $760$ additional confirmed sources for LISA, corresponding to an increase of $7.3\%$ relative to the baseline. For LISA-Taiji, the corresponding numbers are $3229$ additional reported and $2657$ additional confirmed sources, corresponding to an increase of $14.6\%$. These thresholds were determined by iteratively adjusting each $R_{\rm ee}$ until the resulting overall detection rate matched the baseline value, so that the comparison is made at equal overall detection rate. This suggests that the primary effect of the prior is to suppress spurious candidates. Because the prior-imposed catalog is cleaner, this relaxed threshold configuration maintains nearly the same overall detection rate ($84.75\%$) while recovering 658 extra confirmed sources in the low-frequency blocks. This offers a tunable tradeoff between the number of confirmed sources and the detection rate.

\begin{figure*}[t]
\begin{center}
\includegraphics[width=0.95\textwidth]{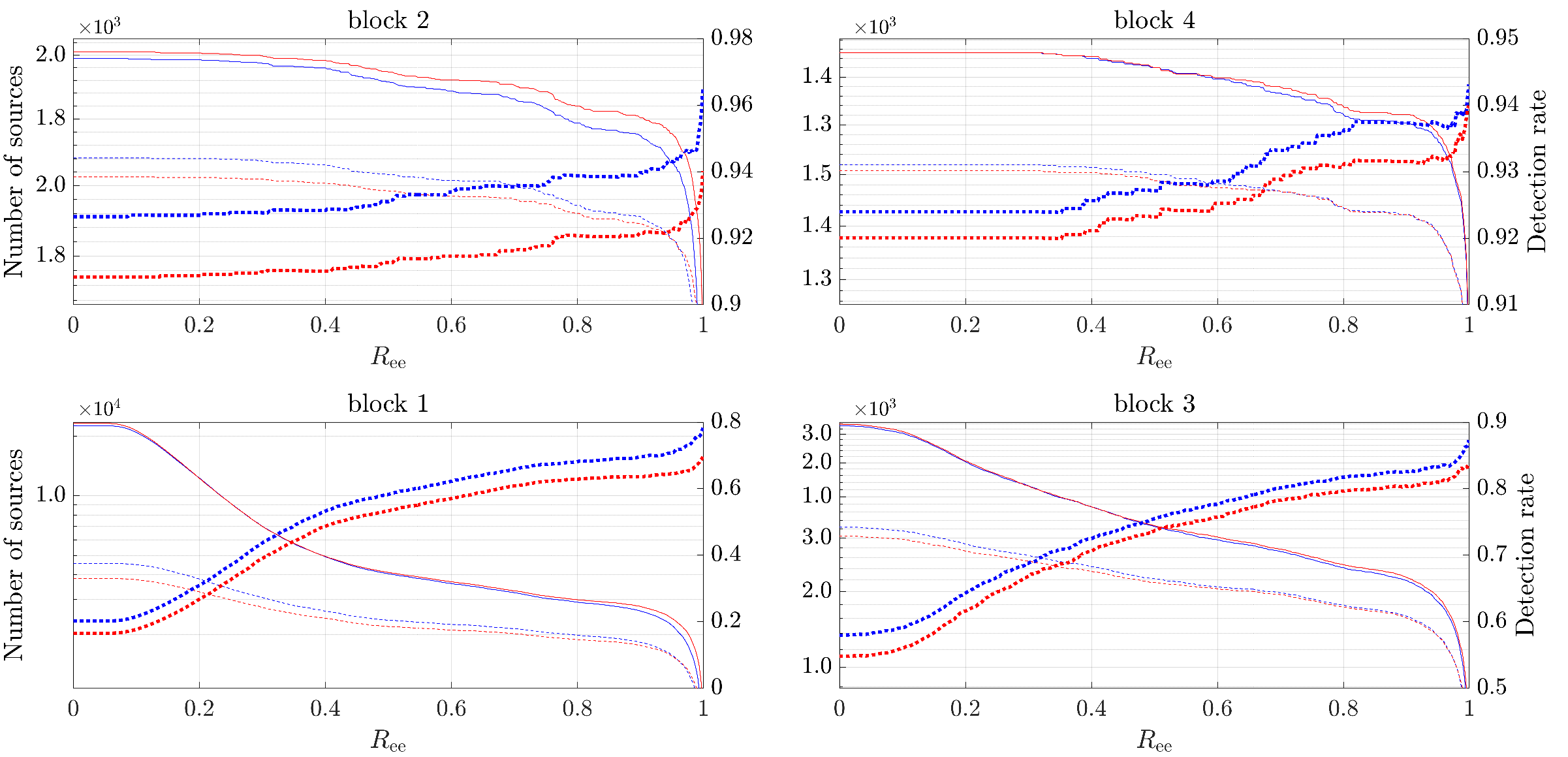}
\caption{Effect of the $R_{\rm ee}$ on the number of reported sources (solid lines), confirmed sources (dashed lines), and detection rate (dotted lines) for the first four frequency--SNR blocks in the single LISA detector configuration obtained from the LDC catalog. Results are shown for block 1-4. Each curve is obtained by scanning the $R_{\rm ee}$ threshold. Red curves correspond to the baseline \fitalgoname{} pipeline, while blue curves incorporate the $\dot{f}$ prior. Left and right vertical axes indicate the number of sources and the detection rate, respectively.}
\label{fig:LISA_LDC_Ree_block}
\end{center}
\end{figure*}

Figure~\ref{fig:LISA_LDC_Ree_block} compares the confirmed-source counts as a function of the $R_{\rm ee}$ threshold for the baseline and prior-enhanced configurations, separately for each frequency block. Several features are noteworthy. First, the curves for different blocks exhibit distinct saturation behaviors: block~1 requires progressively lower $R_{\rm ee}$ values to recover additional sources, consistent with the confusion-dominated nature of this band. Second, the prior's impact is most visible in block~1 and block~2, where the detection curves shift appreciably upward relative to the baseline; in blocks~3 and~4, the gain is more modest, as these bands are already well-resolved even without the prior due to the higher SNR of the sources.

The physical interpretation is straightforward. In low-frequency, crowded bands, the uninformative uniform prior on $\dot f$ allows the likelihood to settle on artificially inflated values, producing false positives that are then filtered out during the $R_{\rm ee}$ validation step. By restricting $\dot f$ to its physically allowed range, the astrophysical prior preemptively suppresses these spurious solutions, leading to a higher fraction of genuine sources surviving the $R_{\rm ee}$ cut. The diminishing returns toward higher SNR blocks are consistent with the fact that those sources are brighter and less vulnerable to $\dot f$-related degeneracies. Taken together, these results demonstrate that the prior's benefit is robust across the full range of $R_{\rm ee}$ thresholds and is precisely where it is needed most.

The recovery statistics in Tables~\ref{tab:snr_ree_LISA} and~\ref{tab:snr_ree_LISA_prior} include both detached ($\dot f > 0$) and semi-detached ($\dot f < 0$) systems. To present the semi-detached results separately, we classify sources by the sign of the estimated $\dot f$. We emphasize that this is an operational classification based on the recovered parameters, and is not necessarily equivalent to the injected system's physical type.

Table~\ref{tab:snr_ree_LISA_prior_semi} summarizes the performance of the single-detector LISA implementation of \fitalgoname\ on the LDC1-4 dataset for semi-detached systems, using the {\it Main} $R_{\rm ee}$ configuration throughout. Compared to the baseline, the prior reduces the number of identified candidates from $8661$ to $8504$ and reported sources from $2796$ to $2564$, indicating effective suppression of spurious candidates. While confirmed sources decrease modestly from $2178$ to $2052$, the overall detection rate improves from $77.90\%$ to $80.03\%$, as the prior preferentially removes false positives rather than genuine signals. Block~1 shows the most pronounced effect, while blocks~4 and~5 exhibit slight gains in both confirmed counts and detection rates, suggesting that the prior is particularly beneficial in the low-frequency regimes where confusion noise is more severe. Overall, the $\dot{f}$ prior enhances catalog purity at the cost of a small reduction in confirmed detections.
\begin{table*}
\begin{center}
\begin{tabular*}{0.8\textwidth}{@{\extracolsep{\fill}}ccccccc}
\toprule
\midrule
\multicolumn{7}{c}{original results}\\
\midrule
 & block~1 & block~2 & block~3 & block~4 & block~5 & Overall\\
\midrule
$R_{\rm ee}$ & $0.9$ & $0.5$ & $0.9$ & $0.5$ & $-1$ & -\\
\midrule
Identified   & $6334$ & $752$ & $1076$ & $230$ & $269$ & $8661$\\
\midrule
Reported     & $1084$ & $742$ & $473$ & $228$ & $269$ & $2796$\\
\midrule
Confirmed    & $738$ & $675$ & $375$ & $193$ & $197$ & $2178$\\
\midrule
Detection rate & $68.08\%$ & $90.97\%$ & $79.28\%$ & $84.65\%$ & $73.23\%$ & $77.90\%$\\
\midrule
\midrule
\multicolumn{7}{c}{piror implemented}\\
\midrule
\midrule
 & block~1 & block~2 & block~3 & block~4 & block~5 & Overall\\
\midrule
$R_{\rm ee}$ & $0.9$ & $0.5$ & $0.9$ & $0.5$ & $-1$ & -\\
\midrule
Identified   & $6229$ & $730$ & $1028$ & $241$ & $276$ & $8504$\\
\midrule
Reported     & $901$ & $720$ & $429$ & $238$ & $276$ & $2564$\\
\midrule
Confirmed    & $624$ & $661$ & $346$ & $204$ & $217$ & $2052$\\
\midrule
Detection rate & $69.26\%$ & $91.81\%$ & $80.65\%$ & $85.71\%$ & $78.62\%$ & $80.03\%$\\
\bottomrule
\end{tabular*}

\caption{Performance of the single-detector LISA implementation of \fitalgoname\ on the LDC1-4 dataset for semi-detached systems ($\dot{f}<0$), using the {\it Main} $R_{\rm ee}$ configuration ($[0.9,0.5,0.9,0.5]$). The upper and lower sections show results without and with the $\dot{f}$ prior, respectively.}
\label{tab:snr_ree_LISA_prior_semi}
\end{center}
\end{table*}

\begin{figure*}
\begin{center}
\includegraphics[width=0.95\textwidth]{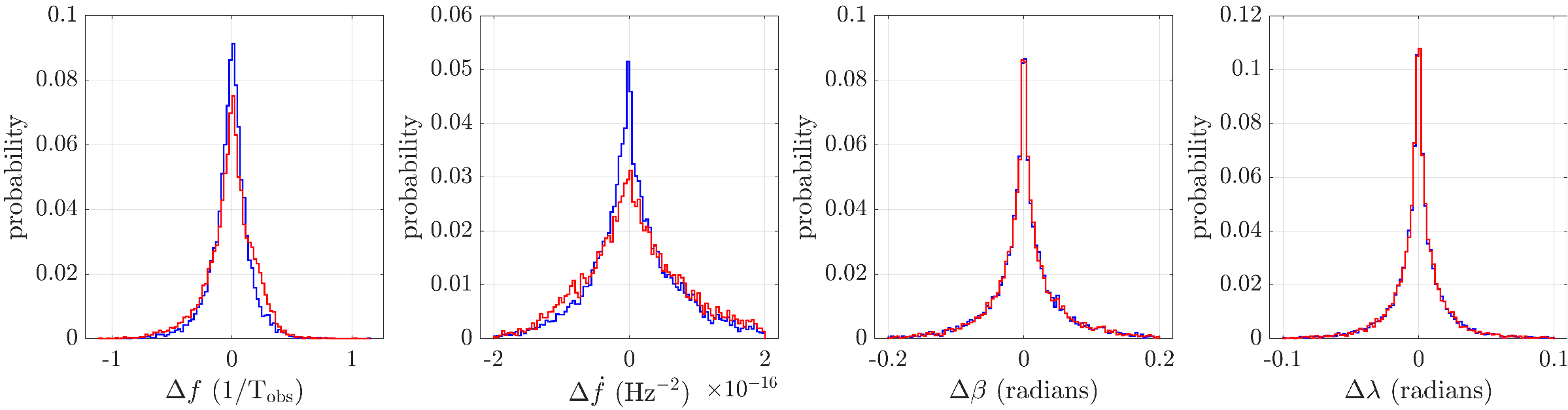}
\caption{Normalized distributions of parameter estimation residuals for the four source parameters $(\Delta f,\, \Delta\dot{f},\, \Delta\beta,\, \Delta\lambda)$ in the single LISA detector configuration, obtained from the LDC catalog. For each source, the residual is computed as the difference between the confirmed sources and the corresponding injected value; the histograms aggregate these residuals over all matched sources. Red and blue histograms correspond to the baseline \fitalgoname{} pipeline and the run incorporating the $\dot{f}$ prior, respectively. The inclusion of the $\dot{f}$ prior yields a notable reduction in the estimation errors of both $f$ and $\dot{f}$, while leaving the sky localization parameters $\beta$ and $\lambda$ largely unaffected.}
\label{fig:Single4ParamError}
\end{center}
\end{figure*}

Figure~\ref{fig:Single4ParamError} compares the distributions of parameter estimation errors for the four phase-evolution parameters between the baseline \fitalgoname configuration and the run incorporating the $\dot f$ prior for a single LISA detector. The overlaid histograms show that imposing the astrophysical prior leads to visibly tighter error distributions for the frequency $f$, and particularly for $\dot f$: the peaks become higher and the wings are suppressed, yielding distributions that are closer to zero bias and less dispersed.

By contrast, the sky localization errors for $\lambda$ and $\beta$ exhibit similar shapes and centroids in both configurations, indicating that the prior primarily improves the phase-evolution parameters ($f$ and $\dot f$) governing chirp evolution rather than the sky location, which is driven by the detector's orbital modulation. This selective improvement is consistent with the prior breaking the $f$--$\dot f$ degeneracy, thereby reducing error propagation into correlated phase-evolution parameters.

\begin{figure*}
\begin{center}
\includegraphics[width=0.85\textwidth]{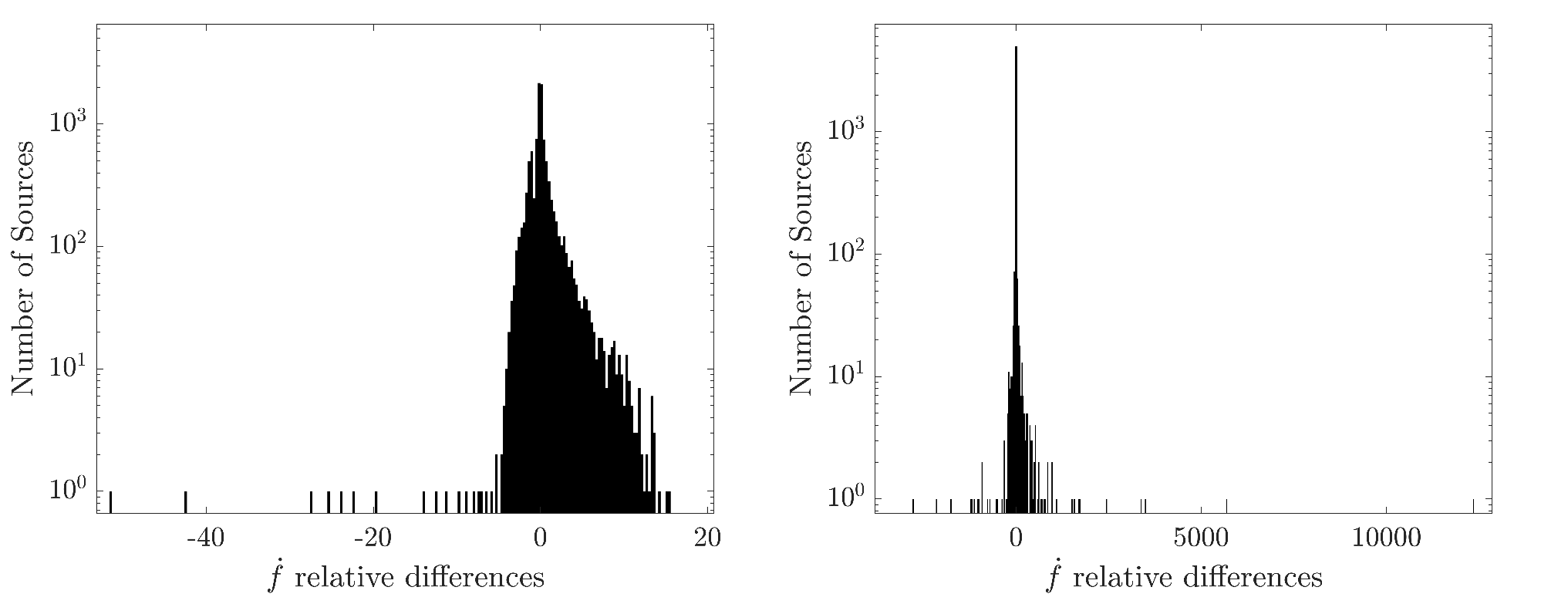}
\caption{Distributions of the relative differences in $\dot{f}$ between the recovered and injected values for confirmed sources, comparing the baseline \fitalgoname{} pipeline (left panel) and the configuration incorporating the $\dot{f}$ prior (right panel). The distributions show the parameter estimation errors between the PSO point estimates and the corresponding injected sources in the LDC catalog.}
\label{fig:fdotRelativeDiff}
\end{center}
\end{figure*}

To isolate the prior's effect on $\dot f$ independently of its absolute scale, Figure~\ref{fig:fdotRelativeDiff} presents the distributions of the relative differences in $\dot{f}$ between the recovered and injected values for both the baseline \fitalgoname{} pipeline and the configuration including the $\dot{f}$ prior. The prior leads to a substantially more concentrated distribution centered near zero: the histogram peak becomes significantly sharper, while the broad asymmetric tails are strongly suppressed. This indicates that the prior effectively constrains the chirp evolution and reduces large deviations in the recovered $\dot{f}$ values.

In the baseline configuration, the relative differences exhibit a much wider spread with extreme outliers extending to both large positive and negative values. Such behavior is consistent with degeneracies in the phase-evolution parameter space, where poorly constrained $\dot{f}$ values propagate into unstable solutions. By incorporating the $\dot{f}$ prior, these degeneracies are broken, yielding a more robust and statistically stable recovery of $\dot f$ across the source population.

\subsection{Network performance}\label{subsec:LISA-Taiji_results}

Network analyses with multiple detectors offer two key advantages: improved sky localization from angular resolution and better parameter estimation through complementary noise realizations. In the case of LISA and Taiji, their orbital configurations yield different modulation patterns for the same source, breaking degeneracies and reducing confusion. The network configuration thus provides a more challenging test bed for the prior, as any improvements must overcome the already enhanced baseline performance.

We evaluate the joint LISA-Taiji detector network configuration using the combined LDC1-4 and Taiji-mod datasets used in previous work~\cite{zhang2022resolving}. The cross-validation configuration mirrors the single detector case, with block dependent $R_{\rm ee}$ thresholds tuned to maintain a balance between completeness and purity.

Table~\ref{tab:snr_ree_LISA_Taiji} lists the baseline performance of the LISA--Taiji network~\cite{zhang2022resolving} under the {\it Main} $R_{\rm ee}$ configuration. The network alone achieves an overall detection rate of $82.53\%$ and confirms $18,\!151$ sources. Performance varies significantly across frequency blocks: block~5 is nearly saturated at $96.78\%$, whereas block~1---the most confusion-dominated band---reaches only $66.05\%$. This block-to-block variation reflects the increasing difficulty of source identification at lower frequencies where source confusion is most severe. 
\begin{table*}
\begin{center}
\begin{tabular*}{0.8\textwidth}{@{\extracolsep{\fill}}ccccccc}
\toprule
 & block~1 & block~2 & block~3 & block~4 & block~5 & Overall\\
\midrule
$R_{\rm ee}$ & $0.9$ & $0.5$ & $0.9$ & $0.5$ & $-1$ & -\\
\midrule
Identified   & $33144$ & $3941$ & $3461$ & $2528$ & $4687$ & $47761$\\
\midrule
Reported     & $8420$ & $3920$ & $2440$ & $2526$ & $4687$ & $21993$\\
\midrule
Confirmed    & $5561$ & $3605$ & $2043$ & $2406$ & $4536$ & $18151$\\
\midrule
Detection rate & $66.05\%$ & $91.96\%$ & $83.73\%$ & $95.25\%$ & $96.78\%$ & $82.53\%$\\
\bottomrule
\end{tabular*}

\caption{Performance of the LISA--Taiji network implementation of \fitalgoname\ on the combined LDC1-4 and Taiji-mod datasets. The results adopt the {\it Main} $R_{\rm ee}$ configuration from \firstpaper. The network configuration achieves consistently high detection efficiencies across all blocks, with an overall detection rate exceeding $82\%$ and a substantial increase in the number of confirmed sources relative to the single-detector case.}
\label{tab:snr_ree_LISA_Taiji}
\end{center}
\end{table*}

Table~\ref{tab:snr_ree_LISA_Taiji_prior} shows the corresponding results after incorporating the astrophysical prior. The number of reported sources in block~5 is 4 fewer than the number of identified sources, because these 4 sources did not satisfy our one-to-one matching criterion. Although the prior leads to $469$ fewer reported sources overall, the number of confirmed sources increases by $1007$, representing a $5.6\%$ gain relative to the baseline confirmed count of 18,151. And the overall detection rate rises from $82.53\%$ to $89.01\%$ under the {\it Main} configuration. The improvement is most pronounced in the low-frequency, low-SNR block, where the detection rate increases from $66.05\%$ to $78.98\%$. Even the challenging $3$--$4$~mHz low-SNR bin improves by approximately four percentage points, while high-SNR regions remain nearly saturated. This suggests that the prior is most effective in confusion-limited regimes where residual signals and spectral leakage from imperfect subtraction can generate spurious maxima in the detection statistic.
\begin{table*}
\begin{center}
\begin{tabular*}{0.8\textwidth}{@{\extracolsep{\fill}}ccccccc}
\toprule
 & block~1 & block~2 & block~3 & block~4 & block~5 & Overall\\
\midrule
$R_{\rm ee}$ & $0.9$ & $0.5$ & $0.9$ & $0.5$ & $-1$ & -\\
\midrule
Identified   & $32615$ & $3919$ & $3448$ & $2527$ & $4742$ & $47251$\\
\midrule
Reported     & $7941$ & $3898$ & $2422$ & $2525$ & $4738$ & $21524$\\
\midrule
Confirmed    & $6272$ & $3706$ & $2123$ & $2433$ & $\deleted{4524} \added{4624}$ & $19158$\\
\midrule
Detection rate & $78.98\%$ & $95.07\%$ & $87.66\%$ & $96.36\%$ & $97.59\%$ & $89.01\%$\\
\midrule
\midrule
\multicolumn{7}{c}{Changing $R_{\rm ee}$ for same detection rate}\\
\midrule
\midrule
 & block~1 & block~2 & block~3 & block~4 & block~5 & Overall\\
\midrule
$R_{\rm ee}$ & $0.425$ & $0.1$ & $0.5$ & $0.1$ & $-1$ & -\\
\midrule
Identified   & $32615$ & $3919$ & $3448$ & $2527$ & $4742$ & $47251$\\
\midrule
Reported     & $11265$ & $3918$ & $2774$ & $2527$ & $4738$ & $25222$\\
\midrule
Confirmed    & $7701$ & $3720$ & $2330$ & $2433$ & $4624$ & $20808$\\
\midrule
Detection rate & $68.36\%$ & $94.95\%$ & $83.99\%$ & $96.28\%$ & $97.59\%$ & $82.50\%$\\
\bottomrule
\end{tabular*}

\caption{Performance of \fitalgoname\ with the astrophysical $f$--$\dot f$ prior for the LISA--Taiji network on the combined LDC1-4 and Taiji-mod datasets. The upper section reproduces the {\it Main} $R_{\rm ee}$ configuration from \firstpaper, where the prior yields an overall detection rate of $89.01\%$ and increases the confirmed sources to $19,\!158$---an improvement relative to the prior-free network baseline in Table~\ref{tab:snr_ree_LISA_Taiji}. The lower section demonstrates that by relaxing $R_{\rm ee}$, the number of confirmed sources can be further increased to $20,\!808$ while maintaining an $82.50\%$ overall detection rate, highlighting the trade-off between prior strength and detection completeness.}

\label{tab:snr_ree_LISA_Taiji_prior}
\end{center}
\end{table*}

Notably, the relative improvement from the prior is larger in the network configuration ($5.6\%$ confirmed-source gain) compared to the single-LISA case ($2\%$ under the same {\it Main} configuration), suggesting that the prior's benefit compounds with improved data quality rather than being absorbed by it. This difference becomes more pronounced when the $R_{\rm ee}$ thresholds are adjusted to equalize the detection rates with and without the prior: the gain increases to $14.6\%$ for the LISA--Taiji network versus $7.3\%$ for the single LISA detector, approximately doubling the relative improvement. When the $R_{\rm ee}$ thresholds are relaxed to equalize the overall detection rate, the prior delivers $2,\!657$ more confirmed sources than the {\it Main} configuration without prior. This indicates that the prior primarily removes inconsistent or unstable reported candidates associated with imperfect subtraction of bright sources. The relaxed configuration adds $3,\!229$ extra reported sources relative to the {\it Main} cuts while maintaining an overall detection rate of $82.50\%$, demonstrating that the prior preserves catalog purity even under more aggressive completeness settings.

\section{Conclusion}\label{sec:conclusion}
We have presented an astrophysically motivated joint prior on $(f,\dot f)$ for Galactic double white dwarfs and incorporated it into the \fitalgoname data analysis pipeline via a Tukey window regularization of the $\mathcal{F}$-statistic along the $\dot{f}$ direction. Maintaining the same overall detection rate, the prior increases the number of confirmed detections by approximately $7.3\%$ for a single LISA interferometer and $14.6\%$ for the LISA--Taiji network. These gains are accompanied by improved parameter recovery for $f$ and $\dot{f}$, cleaner subtraction of bright binaries, and enhanced recovery of low-SNR sources in confusion-limited frequency bands.

The improvement in parameter estimation is reflected in tighter and more centrally peaked error distributions for $f$ and $\dot{f}$, with fewer large outliers. This behavior is consistent with the prior reducing the degeneracy between these two phase-evolution parameters, particularly in the low-SNR regime. By contrast, sky localization errors remain largely unchanged, indicating that the prior primarily constrains intrinsic waveform evolution rather than parameters determined by the detector response.

From an operational perspective, the prior enables the use of more relaxed $R_{\rm ee}$ thresholds to improve completeness while maintaining comparable catalog purity. This reduces the number of spurious candidates introduced at low SNR and can help limit the manual effort required for catalog validation. In addition, by suppressing residual-induced artifacts associated with imperfect subtraction of bright sources, the prior supports more stable iterative subtraction strategies in which source recovery and model refinement are performed jointly.

The application and interpretation of our frequency drift prior come with inherent limitations tied to the underlying assumptions and calibration framework. The prior is constructed from the LDC catalog, representing one specific realization of Galactic double white‑dwarf populations. Moreover, the current prior worked both on detached systems and semi-detached systems. But with the associated mass‑transfer effects for semi-detached systems, more sophisticated model is needed. A further caveat lies in the idealized noise and response assumed in the detection pipeline; in practice, calibration uncertainties could diminish the prior's effectiveness.

On the implementation side, the use of a fixed $\alpha = 0.9$ for the Tukey window offers a practical baseline, but adaptive schemes that tune $\alpha$ based on data characteristics could offer further gains. Finally, our statistical assessment relies on the assumption of independent observations, which may be adequate for well‑separated sources but demands careful scrutiny in regions of high confusion.

Future work will focus on modeling the analytical prior form for the semi-detached systems, where mass transfer introduces additional complexity beyond the simplified treatment used here. Another direction is to explore whether the dimensionless parameter $\dot f_{\rm Ratio}$ introduced in this work can serve directly as a search parameter in \fitalgoname to further improve performance. We also plan to investigate adaptive schemes that update the prior as high-confidence detections accumulate, as well as testing against future binary population synthesis releases that include interacting binaries and metallicity gradients. Together, these efforts will help ensure that astrophysical priors remain a robust and transparent tool for maximizing the scientific return from the Galactic DWD foreground.

\begin{acknowledgments}
This study was supported by the National Key Research and Development Program of China (Grant No. 2021YFC2203003 and No.2023YFC2206701), the National Natural Science Foundation of China (Grants No. 12475056, No. 12247101), the Fundamental Research Funds for the Central Universities(Grant No. lzujbky-2025-jdzx07) , the Natural Science Foundation of Gansu Province (No. 22JR5RA389, No. 25JRRA799).

\end{acknowledgments}

\bibliography{refs}
\end{document}